\documentclass{ws-mplb}

\begin{document}

\markboth{Dan-Dan Su, Xi Dai, Ning-Hua Tong}{Local Entanglement Entropy at the Mott Metal-Insulator Transition in Infinite Dimension}

%
\catchline{}{}{}{}{}
%

\title{Local Entanglement Entropy at the Mott Metal-Insulator Transition\\
 in Infinite Dimensions}

\author{Dan-Dan Su}

\address{Department of Physics, Renmin University of China,\\
Beijing 100872, P. R. China
}

\author{Xi Dai}

\address{Beijing National Laboratory for
Condensed Matter Physics, Institute of Physics,\\
Chinese Academy of Sciences, Beijing 100190, China}

\author{Ning-Hua Tong}

\address{Department of Physics, Renmin University of China,\\
Beijing 100872, P. R. China\\
nhtong@ruc.edu.cn}

\maketitle

\begin{history}
\received{(Day Month Year)}
\revised{(Day Month Year)}
\end{history}

\begin{abstract}
We study the critical behavior of the single-site
entanglement entropy $S$ at the Mott metal-insulator transition in
infinite-dimensional Hubbard model. For this model, the entanglement
between a single site and rest of the lattice can be evaluated
exactly, using the dynamical mean-field theory (DMFT). Both the
numerical solution using exact diagonalization and the analytical
one using two-site DMFT gives $S -S_c \propto \alpha \log_{2}\left[
(1/2-D_c)/ D_c \right] (U-U_c)$, with $D_c$ the double occupancy at
$U_c$ and $\alpha < 0$ being different on two sides of the
transition.
\end{abstract}

\keywords{Hubbard model; Metal-insulator transitions; Entanglement entropy.} 

\section{Introduction}

    The concept of quantum entanglement plays a key role in the
field of quantum information manipulation and
processing\cite{Vincenzo00247}. It describes the inseparability
between parts of a given system in a given state. In the past
decade, the close relation between quantum entanglement and the
quantum correlation in many-body systems are
discussed\cite{Amico08517}, especially in the context of condensed
matter physics\cite{Thunstrom12}, cold
atoms\cite{Buonsante07110601}, and quantum
chemistry\cite{Babamoradi10}, etc.

    One of the most interesting ideas is to employ the entanglement
entropy as an indicator of quantum phase transition
(QPT)\cite{Sachdev1} in both spin systems\cite{Osborne0232110,Melko10100409,Kallin11165134,Hastings10157201,Singh1275106,Verstraete0487201,Popp0542306}
and the interacting fermion
systems\cite{Zanardi0242101,Wang0422301,Gu0486402,Johannesson07935,Larsson05196406,Larsson0642320,Chan08345217}.
For those quantum phase transitions that cannot be described by
Landau's symmetry breaking paradigm, there is no well defined local
order parameters. In such cases, the entanglement entropy may be a
really useful concept for characterizing the quantum phase.

   One of such QPT is the Mott metal-insulator transition, where a
 many-body system transits from a metallic state into an insulator,
when the interaction strength between particles exceeds a critical
value\cite{Mott49416}. Experimentally, Mott transition (MT) has been
widely studied both in strongly correlated electron systems such as
$V_2O_3$\cite{McWhan703734} and in cold atom
systems\cite{Greiner0239}. In the most strict sense of the Mott
transition, no symmetry breaking occurs at the transition and hence
it belongs to the type outside the Landau's paradigm. Since 1960's,
Mott transition has been one of the key issues in condensed matter
physics. Theoretically, intensive studies based on the Hubbard-type
models have been done in the past decades\cite{Gebhard1}. The study
of MT in terms of the quantum entanglement, especially using the
entanglement entropy as a measure, appears for
one-dimensional\cite{Gu0486402,Larsson05196406,Larsson0642320,Chan08345217},
two-dimensional\cite{Wang0422301}, and infinite
dimensional\cite{Byczuk1287004} Hubbard-like models.

For fermionic lattice models, a connection has been established
rigorously between the singularity of single-site entanglement
entropy $S$ and the order of QPT, under certain
conditions\cite{Larsson0642320}. This connection states that the
discontinuity in the (k - 1)-th order derivative of S gives a k-th
order QPT. For one dimensional Hubbard model, $S$ reaches a maximum
at $U_c=0$ where MT occurs\cite{Gu0486402}, due to equal population
of all the local bases at the transition
point\cite{Johannesson07935}. In two dimensions, study on finite
size system does not disclose any singularity at
$U_c$\cite{Wang0422301}. In this paper, we focus on the single-site
entanglement entropy $S$ near the MT in the fermionic Hubbard model
in infinite spatial dimensions. In this limit, the spatial
fluctuations of electrons are suppressed while the local quantum
fluctuation remains. The Hubbard model can be solved exactly in this
limit using the dynamical mean-field theory
(DMFT)\cite{Metzner89324,Georges9613}. Recently, DMFT is used to
evaluate the relative entropy\cite{Gottlieb07815} as a measure of
correlation for the Hubbard model as well as for a series of
transition metal oxides\cite{Byczuk1287004}. For the half-filled
Hubbard model, at low temperature, the Fermi-liquid state in small
$U$ regime is separated from the Mott insulator state in large $U$
regime by a special second-order
QPT\cite{Georges9613,Bulla99136,Tong01235109}. Although this Mott
transition in large spatial dimensions has received considerable
attention in the past years, no analysis has been carried out for
the critical behavior of its entanglement properties. We expect that
such analysis can help to deepen our understanding of the MT in
general.

The structure of this paper is as follows. In Section $2$, we
introduce the single-site entanglement entropy $S$ and the method
that we used to evaluate it for the Hubbard model, {\it i.e.}, the
DMFT with exact diagonalization and the two-site DMFT. In Section
$3$, we present numerical as well as analytical results for $S$ near
the MT. In Section $4$ we end with a brief summary.

\section{Model and Method}

   The Hamiltonian of the Hubbard model reads $(\hbar=1)$
\begin{eqnarray}\label{eq:1}
H &=& - \sum_{i,j} t_{ij} c_{i}^{\dagger} c_{j} + U \sum_{i} n_{i
\uparrow} n_{i \downarrow} - \mu \sum_{i \sigma} n_{i \sigma}.
\end{eqnarray}
Here, $t_{ij}$ is the hopping matrix element and $U$ is the on-site
repulsion of electrons with opposite spin. $c_{i \sigma}$ and $c_{i
\sigma}^{\dagger}$ are annihilation and creation operators of the
electron on site $i$ with spin $\sigma$, respectively. $\mu$ is the
chemical potential.

To study the local entanglement entropy, we divide the whole lattice
into two parts, subsystem (A) (a single site $i$) and the
environment (B) (the rest part of the lattice). For a given quantum
state $|\psi \rangle$ of the whole system, the reduced density
matrix of the subsystem is
\begin{equation}
\hat{\rho}_\mathrm{A}=\mathrm{Tr}_\mathrm{B}|\psi\rangle\langle\psi|
.
\end{equation}
The  bipartite entanglement entropy between the subsystem and the
environment is defined as (setting $k=1$)
\begin{equation}
S=-\mathrm{Tr}(\hat{\rho}_\mathrm{A}\log_{2}\hat{\rho}_\mathrm{A})
 =-\mathrm{Tr}(\hat{\rho}_\mathrm{B}\log_{2}\hat{\rho}_\mathrm{B}) .
\end{equation}
From Eq.(2), one gets $\langle \psi |O_{A}| \psi \rangle =
\mathrm{Tr} (\rho_{A}O_{A})$ for any given operator $O_{A}$ of the
system. Taking $O_{i}=1$, $n_{i \uparrow}$, $n_{i \downarrow}$, and
$n_{i \uparrow}n_{i \downarrow}$ for site $i$, one gets 4 equations
about the diagonal elements of $\rho_{i}$ under the basis set $(|0
\rangle, |\uparrow \rangle, |\downarrow \rangle, |\uparrow
\downarrow \rangle)$. Here we study the symmetry unbroken ground
state of Hubbard model $ |\psi \rangle$. The off-diagonal elements
are all zero due to the $U(1)$ and $SU(2)$ symmetries of Hubbard
model, i.e., $\langle \psi | c_{i \sigma}^{\dagger} |\psi \rangle =
\langle  \psi | c_{i \uparrow}^{\dagger} c_{i \downarrow}^{\dagger}
|\psi \rangle = \langle \psi | c_{i \uparrow}^{\dagger} c_{i
\downarrow} |\psi \rangle = 0$. We therefore obtain
\begin{equation}
\rho_\mathrm{A} = \left(
\begin{array}{cccc}
 \langle(1-n_\uparrow)(1-n_\downarrow)\rangle & 0 & 0 & 0 \\
 0 & \langle n_\uparrow(1-n_\downarrow)\rangle & 0 & 0 \\
 0 & 0 & \langle(1-n_\uparrow)n_\downarrow\rangle & 0 \\
 0 & 0 & 0 & \langle n_\uparrow n_\downarrow\rangle
\end{array}
\right).
\end{equation}
The averages are with respect to the ground state of Hubbard model,
and translation invariance is assumed here. For a half-filled
lattice in the paramagnetic phase where the Mott transition occurs,
one has $\langle n_{\uparrow} \rangle = \langle
n_{\downarrow}\rangle = 1/2$. The entanglement entropy thus
reads~\cite{Zanardi0242101}
\begin{equation}
S=-2 \left(\frac{1}{2}-D \right) \log_{2} \left(\frac{1}{2}- D
\right)-2D \log_{2} D .
\end{equation}
Here $D \equiv \langle n_{\uparrow} n_{\downarrow} \rangle$ is the
expectation value of the double occupancy. For the the ground state
of the Hubbard model in infinite spatial dimensions, this quantity
can be readily evaluated from the converged self-consistent solution
of DMFT.

The DMFT is a well-developed theory for treating the Hubbard-type
strongly correlated models\cite{Georges9613}. In DMFT, the Hubbard
model is first mapped into an effective Anderson impurity model,
\begin{equation}
   H_{imp} = \sum_{k \sigma} \left[ \epsilon_{k} a_{k \sigma}^{\dagger}
   a_{k \sigma} + V_{k} \left( a_{k \sigma}^{\dagger} c_{\sigma} + c_{\sigma}^{\dagger} a_{k \sigma}
   \right) \right] + U c_{\uparrow}^{\dagger} c_{\uparrow} c_{\downarrow}^{\dagger}
   c_{\downarrow} -\mu \sum_{\sigma}c_{\sigma}^{\dagger} c_{\sigma}.
\end{equation}
Here $a_{k\sigma}$ is the annihilation operator of bath site $k$,
and the the parameters of the electron bath $\{\epsilon_k, V_k\}$
determine the dynamical "Weiss field"
\begin{equation}
 \mathcal{G}_{0}^{-1}(i \omega_n) = i \omega_n + \mu - \sum_{k}
 \frac{V_{k}^2}{i \omega_n - \epsilon_k}.
\end{equation}
The impurity model is then solved to generate the impurity Green's
function $G(i\omega_n)$ on the Matsubara frequency axis. Finally,
through the self-consistent equation
\begin{equation}
  G(i\omega_n) = \int_{-\infty}^{\infty} \frac{\rho_{0}(\epsilon)}{i \omega_n + \mu -\epsilon - \Sigma(i
  \omega_n)} d\epsilon
\end{equation}
together with $\Sigma(i \omega_n) = \mathcal{G}^{-1}_{0}(i \omega_n)
- G^{-1}(i \omega_n)$, a new "Weiss field"
$\mathcal{G}_{0}^{-1}(i\omega_n)$ can be obtained and used to update
the bath parameters $\{ \epsilon_k, V_k \}$. This process iterates
and the converged solution of the impurity self-energy will be taken
as the local self-energy of the lattice model, i.e., $\Sigma_{ij}(i
\omega_n) = \Sigma_{imp}(i \omega_n)\delta_{ij}$.

We adopt the semi-circular density of states
\begin{equation}
    \rho_{0}(\epsilon) = \frac{2}{\pi W^{2}} \sqrt{W^2 - \epsilon^2}.
\end{equation}
$W=1$ is set as the energy unit. Eq.(9) is the density of states of
free electrons on the Bethe lattice with infinite coordination
number. It is widely used in the study of the MT because it
simplifies the self-consistent equation while keeps the qualitative
physics intact. We first solve DMFT equations using the exact
diagonalization (ED) method of Caferral {\it et
al.}\cite{Caffarel941545}. Then, we resort to the two-site
DMFT\cite{Potthoff01165114} for analytical results. At half filling,
this theory is reduced to the linearized DMFT\cite{Bulla00257}. It
is shown that it can produce rather accurate $U_c$ as well as
physical quantities near the critical point.

   In the following, we present our results for $D$ and $S$ as a
function of $U$, both from DMFT with ED and from the analytical
two-site DMFT formulism.

\section{Results and Discussions}

\subsection{Exact diagonalization results}
\begin{figure}[th]
\begin{center}
\includegraphics[width=8cm,angle=-90]{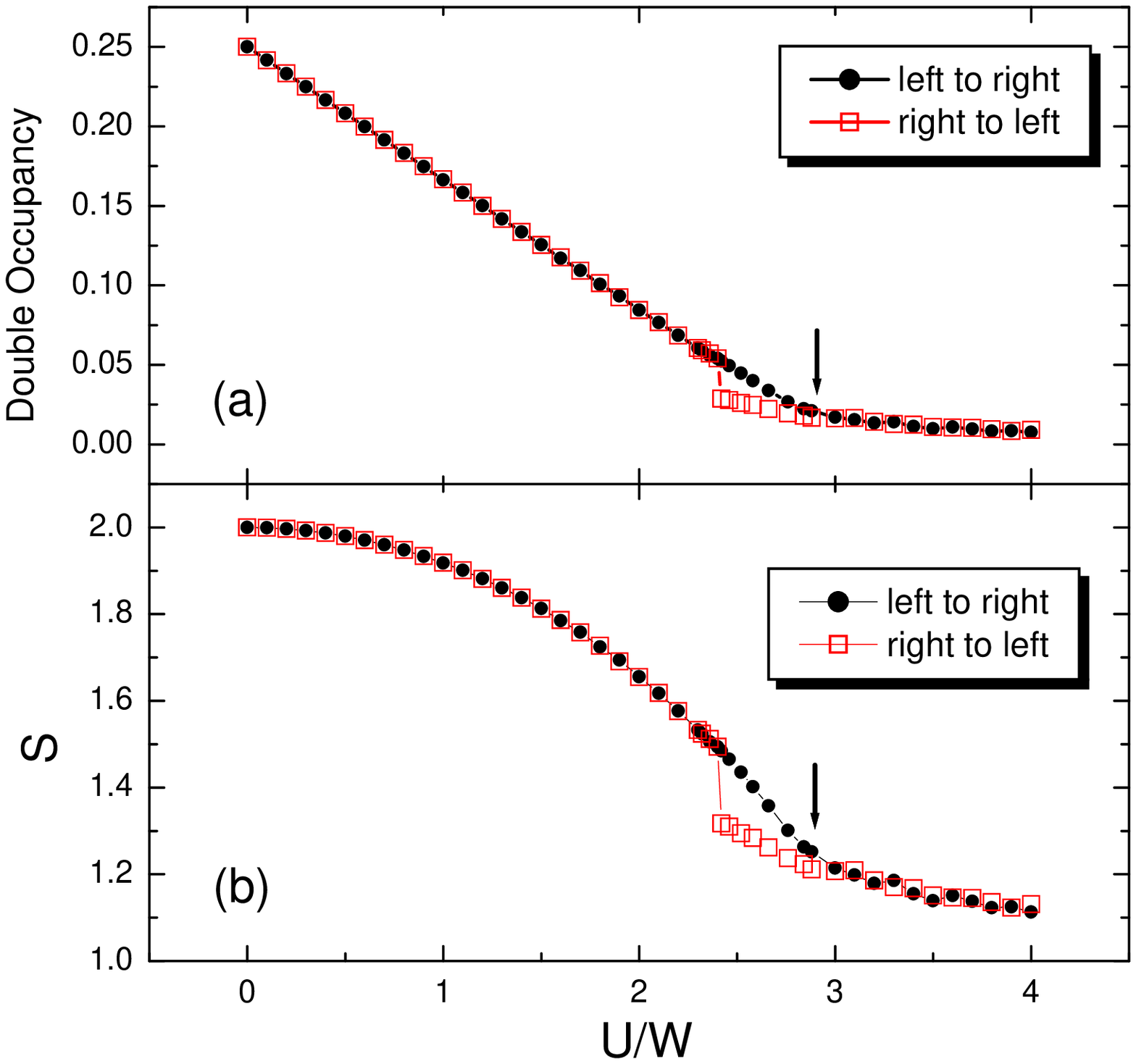}
\end{center}
\vspace*{8pt}
\caption{(a) Double occupancy $D$ and (b) entanglement entropy $S$
as functions of $U$, obtained from ED calculation with $N_{s}=6$.
Black solid dots are obtained by scanning from small $U$ to large
$U$, and red empty squares are obtained by scanning from large $U$
to small $U$. Arrows mark the transition point $U_{c2}$. Other
parameters are $\mu$=U/2 and $T=10^{-4}W$. \label{fig1}}
\end{figure}

It has been known\cite{Georges9613,Tong01235109} that for
finite temperatures $0 < T < T_c$, the metallic state for small $U$
is separated from the insulating-like state for large $U$ by a
finite regime $U_{c1}(T) < U < U_{c2}(T)$, in which the metal and
the insulator phases coexist. The true first order phase transition
occurs at $U_c(T)$ ( $U_{c1}(T) < U_c(T) < U_{c2}(T)$ ) where the
free energies of the metallic and of the insulating solutions
coincide. At zero temperature, $U_{c1} \approx
2.38W$\cite{Tong01235109} and $U_c = U_{c2} \approx
2.94W$\cite{Bulla99136}. The Mott transition becomes a special
second-order phase transition at $U_{c2}(T=0)$: $\partial E_{g}
/\partial U$ is continuous but the meta-stable solution of insulator
extends from large U regime into the regime $U_{c1}(0) < U <
U_{c2}(0)$.

In Fig.1, we show the double occupancy $D$ and entanglement entropy
$S$ as functions of $U$, obtained from DMFT with ED for a very low
temperature $T=10^{-4}W$, which is practically same as zero
temperature. As shown in Fig.1(a), the double occupancy $D$
decreases linearly as $U$ increases up to the critical point
$U_{c2}$. Near $U_{c2}$, coexistence of two solutions and hysteresis
in $U$-scanning are observed. From the solution of $D$ obtained by
scanning from small $U$ to large $U$ (solid circles in Fig.1(a)), we
can identify a linear form in the $U < U_{c2}$ regime after a
non-singular term $D_c$ is subtracted,
\begin{equation}
    D - D_c \propto U_{c2} - U,  \,\,\, \,\,\, (U < U_{c2}).
\end{equation}
The other solution of $D$ (empty squares in Fig.1(a)) is obtained by
scanning from large $U$ to small $U$. It extends to $U_{c1}$ and
recovers the first solution through a finite jump. From the
continuation of $D$ at $U_{c2}$ in the second solution, we can infer
that in the $U
> U_{c2}$ regime, a linear behavior with a much smaller slope must
hold. Therefore, we can summarize $D(U)$ as
\begin{equation}
    D = D_c  + \alpha ( U - U_{c2} ),
\end{equation}
with $\alpha < 0$ and having different values on two sides of the
phase transition. This is consistent with the scenario of the
special second-order MT\cite{Tong01235109} in infinite dimensions:
$\partial^2 E/ \partial U^2 \sim \partial D/
\partial U$ is discontinuous. Our ED calculation gives $U_{c1} \approx 2.4W$
 and $U_{c2} \approx 2.9W$, in agreement
with previous results of ED and projected self-consistent
technique~\cite{Georges2}. There is a small but finite double
occupancy $D_c \approx 0.02$ at $U=U_{c2}$. This reflects that even
at critical point and in the insulating state, there is residual
local charge fluctuations. This has important consequence for the
entanglement entropy in the insulating state. See below.

In Fig.1(b), the entanglement entropy $S$ is shown. At $U=0$, $S=2$
comes from the equal population of electrons on the four local
states. $S$ decreases monotonously with $U$. After a inflection
point at $U=U_{c2}$, it continues to decrease towards its
strong-coupling limit $S_{\infty} = 1$, which comes from the spin
two-fold degeneracy in the paramagnetic insulator phase. Being
consistent with $D$, two solutions coexist in the regime $U_{c1} < U
< U_{c2}$ and a hysteresis is observed. What is interesting is the
nonzero critical value of the entanglement entropy $S_{c} -
S_{\infty} \approx 0.24$. This reflects the residual entanglement
between a single site and the others at the critical point and in
the insulating phase. The critical behavior of $S$ may be easily
obtained from Eq.(8) and (5):
\begin{equation}
    S-S_{c} = 2\alpha \log_{2}\left(\frac{1/2 - D_{c}}{D_c} \right)
    \left(U-U_{c} \right).
\end{equation}
Here $\alpha < 0$ and it has different values on two sides of MT.
Due to the finite $D_{c}$, $S$ is not singular at $U=U_{c2}$ but its
derivative is discontinuous. This is the main result of this paper.

Here is the big difference between the static mean-field theory and
the DMFT. For static mean-field theories such as the Weiss
mean-field theory, the entanglement will become zero as soon as the
system enters the long-range ordered phase, while it stays as a
constant $S=1$ in the paramagnetic phase, arising from spin two-fold
degeneracy. For DMFT, although the short range spin-correlation is
not taken into account, the local temporal fluctuations do generate
entanglement between one site and the others. Therefore, we conclude
that for the Hubbard model, the local charge fluctuations in the
insulating phase is associated with a nonzero local entanglement
between one site and the others.

Here we did not quest for the highest precision in our ED
calculation. Near $U_{c2}$, the critical slowing down in solving
DMFT equations prevents ED from obtaining rigorous conclusion. To
get a more explicit solution, we resort to the two-site DMFT, which
can give analytical results in the critical regime.

%
\subsection{Two-site DMFT result}

In the two-site DMFT, the lattice Hamiltonian is first mapped into
an Anderson impurity model with one bath site,
\begin{equation}
H_{imp}= U c_{\uparrow}^{\dagger} c_{\uparrow}
c_{\downarrow}^{\dagger} c_{\downarrow}
 - \frac{U}{2} \sum_{\sigma} c_{\sigma}^{\dagger} c_{\sigma}
 + \sum_{\sigma} \left[ V \left( c^{\dagger}_{\sigma} a_{\sigma}+a^{\dagger}_{\sigma} c_{\sigma} \right)
                     + \epsilon a^{\dagger}_{\sigma} a_{\sigma} \right].
\end{equation}
The bath parameters $\epsilon$ and $V$ are determined
by~\cite{Potthoff01165114}
\begin{eqnarray}
   n_{imp} = n_{lat},  \nonumber \\
   V^2 = z M_{2}^{(0)}.
\end{eqnarray}
Here $n_{imp}$ and $n_{lat}$ are electron density for the impurity
site and for the lattice site, respectively. $z$ is the
quasi-particle weight and $M_{2}^{(0)}$ is the second order moment
of the free density of states, $M_{2}^{(0)} \equiv
\int_{-\infty}^{\infty} \rho_{0}(\omega) \omega^2 d\omega$. For the
semicircular density of states in Eq.(9), $M_{2}^{(0)}=W^2/4$. For
half-filling, the particle-hole symmetry in our model guarantees
that $\epsilon=0$. $V$ is to be fixed by the second equation of
Eq.(14).

Eq.(13) and (14) can be solved numerically. The $n_{imp}$ and
$n_{lat}$ are obtained from numerical diagonalization of $H_{imp}$
and from the lattice Green's function, respectively. $z$ is obtained
from the weight of the quasiparticle poles of the impurity Green's
function. The numerical results for $D$ and $S$ are shown in Fig.2.

The two-site DMFT can also be solved analytically. For this purpose,
we first solve the retarded Green's function of the impurity model
at zero temperature ($\eta=0^{+}$),
\begin{equation}
G_\sigma(\omega+ i \eta) = \sum_{n} \left[ \frac{|\langle
G|c_\sigma^\dagger|n\rangle|^2}{\omega + i \eta + E_{n} - E_{g}} +
\frac{|\langle n|c_\sigma^\dagger|G\rangle|^2}{\omega+ i
\eta+E_{g}-E_{n}}\right] .
\end{equation}
For the particle-hole symmetric case, the eigen values and eigen states of $H_{imp}$ can be solved
analytically. The Green's function has four poles. Two of them near $\pm
\frac{1}{2}U $ are for Hubbard bands, and the other two poles near $\pm
V^2/U$ are precursors of the coherent Kondo resonance. To calculate
$z$, we take the weights of the latter two poles and expand $z$ in
terms of powers of $V/U$,
\begin{equation}
  z = \frac{36 V^2}{U^2} -\frac{1584 V^4}{U^4} + \ldots.
\end{equation}
This expression together with Eq.(14) gives the solution for $V$ as
\begin{equation}
   V = \frac{U}{2\sqrt{11}}\sqrt{1-U^2/U_{c}^2} \,\,\,\,\,\,\,\,\, (U <
   U_{c}),
\end{equation}
and $V=0$ for $U > U_c$. Here $U_c=6\sqrt{M^{(0)}_2}$. For the Bethe
lattice, $U_c = 3W$ is very close the the ED results $U_{c2} \approx
2.9W$. For $U < U_c$, $V>0$ and the system is in the metallic phase.
For $U > U_c$, $V=0$ and the system is in the insulating phase. Note
that this expression differs from that of
Potthoff\cite{Potthoff01165114} in the prefactor, due to different
ways of calculating $z$\cite{Explain1}.
\begin{figure}[th]
\begin{center}
\includegraphics[width=8cm,angle=-90]{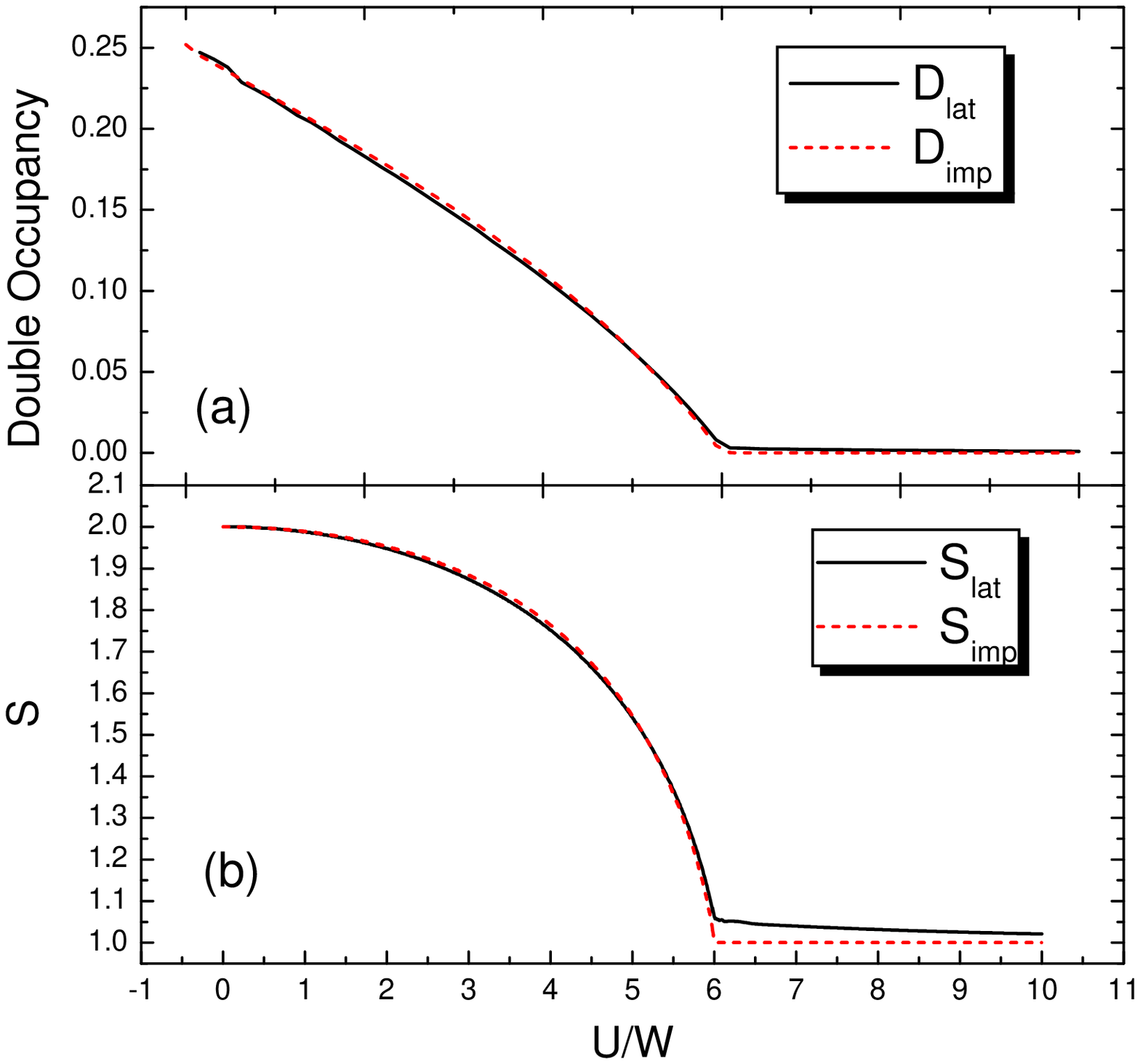}
\end{center}
\vspace*{8pt}
   \caption{Double occupancy $D$ (a) and entanglement entropy $S$ (b)
as functions of $U$, obtained from numerical solution to the
two-site DMFT equations. The solid lines are results for lattice
model, and the dash-dot lines are for impurity model.} \label{fig2}
\end{figure}

Exact solution of $H_{imp}$ gives the analytical expression for the
double occupancy as
\begin{equation}
  D_{imp} \equiv \langle n_\uparrow n_\downarrow\rangle = \frac{\left(U-\sqrt{U^2+64
    V^2}\right)^2}{4 \left(U^2+64 V^2-U \sqrt{U^2+64 V^2}\right)} .
\end{equation}
Combining it with Eq.(17), one obtains the critical expression for
$D$ as,
\begin{equation}
   D_{imp} = \frac{4}{11} \left(1-U/U_{c} \right) \,\,\,\,\,\,\,\,\, (U < U_{c}),
\end{equation}
and $D_{imp}=0$ for $U > U_c$. For $S$ we have
\begin{equation}
    S_{imp}= -\frac{8}{11} \left( 1 - \frac{U}{U_c} \right) \log_{2}\left(1 - \frac{U}{U_c} \right)  \,\,\,\,\,\,\,\,\, (U <
    U_{c}),
\end{equation}
and $S_{imp} = 1$ for $U > U_c$.

However, it is noted that $S_{imp}$ in Eq.(20) is not the correct
results. Although the two-site DMFT equation requires $n_{imp} =
n_{lat}$, the double occupancy $D_{imp}$ does not equal to its
lattice counterpart $D_{lat}$, due to the fact that the two-site
DMFT is an approximation to the full DMFT. In Fig.2(a), we plot
$D_{imp}$ and $D_{lat}$ as functions of $U$. $D_{lat}$ is calculated
using the lattice Green's function\cite{Georges9613}. It is seen
that for $U < U_c$, both $D_{imp}$ and $D_{lat}$ are linear near
$U_c$. For $U > U_c$, although $D_{imp} =0$, $D_{lat}$ is small but
finite, with a finite slope at $U = U_{c}$, consistent with ED
results.

Correspondingly, $S_{imp}$ calculated from $D_{imp}$ becomes $1$
immediately after $U > U_{c}$, while $S_{lat}$ calculated from
$D_{lat}$ is not singular at $U_{c}$. It has a linear behavior at
$U_c$ but with a much smaller slope than that on the $U<U_c$ side.
This agrees with the ED results. Therefore, we believe that the
$S_{lat}$ is an improvement over $S_{imp}$ which ignores the finite
$D$ at $U_c$. This is expected because $S_{lat}$ takes into account
the lattice information more elaborately than $S_{imp}$ through the
lattice Green's function. In conclusion, both the $S_{lat}$ from
two-site DMFT calculation and that of DMFT with ED give the critical
behavior of the single-site entanglement entropy as described by
Eq.(12).

Our conclusion Eq.(12) is exact for the Hubbard model in infinite
dimensions. Being different from the one-dimensional\cite{Gu0486402}
and two-dimensional cases\cite{Wang0422301}, our study shows that MT
in infinite dimensions fulfills the rigorous theorem by Larsson and
Johannesson, which states that the discontinuity in the $(k-1)$-th
order derivative of $S$ gives a $k$-th order
QPT\cite{Larsson0642320}, with $k=2$ in this case. Besides the
single-site entanglement entropy, two-site von Neumann entropy has
been studied at the quantum phase transition\cite{Johannesson07935}.
For Hubbard model, it can be calculated using the cluster extension
of DMFT\cite{Georges9613}.

\section{Summary}

    To conclude, we have analyzed the critical behavior of the single-site
entanglement entropy $S$ at the MT in infinite spatial dimensions,
using DMFT with ED and the two-site DMFT. Even in the insulating
phase, temporal local fluctuations generate nonzero $S$ besides the
contribution from spin two-fold degeneracy. The critical behavior of
$S$ at the MT is not singular due to finite double occupancy at
$U_{c}$, but $\partial S/ \partial U$ is discontinuous at $U_c$,
being consistent with the second order scenario of MT and the
Larsson-Johannesson theorem\cite{Larsson0642320}.

   This work is supported by National Program on Key Basic Research Project
(973 Program) under grant 2012CB921704, and by the NSFC under grant
number 11074302.


\begin{thebibliography}{0}
\bibitem{Vincenzo00247} D. P. Di Vincenzo and C. Bennet,
\newblock Nature (London) {\bf 404}, 247 (2000).

\bibitem{Amico08517} L. Amico, R. Fazio, A. Osterloh, and V. Vedral,
\newblock Rev. Mod. Phys. {\bf 80}, 517 (2008).

\bibitem{Thunstrom12} P. Thunstr\"{o}m, I. Di Marco, and O. Eriksson,
\newblock arXiv:1202.3975.

\bibitem{Buonsante07110601} P. Buonsante 1 and A. Vezzani, Phys. Rev. Lett.
\newblock {\bf 98}, 110601 (2007).

\bibitem{Babamoradi10} M. Babamoradi, M. Heidari Saani, and M. A. Vesaghi,
\newblock arXiv:1007.3112;
N. M. Tubman and J. McMinis,
\newblock arXiv:1204.4731;
K. Boguslawski, P. Tecmer, \"{O}rs Legeza, and M. Reiher
\newblock arXiv:1208.6586.

\bibitem{Sachdev1} S. Sachdev,
\newblock {\em Quantum Phase Transitions},
\newblock Cambridge University Press, 1999.

\bibitem{Osborne0232110} T. J. Osborne and M. A. Nielsen,
\newblock Phys. Rev. A {\bf 66}, 032110 (2002).

\bibitem{Melko10100409} R. G. Melko, A. B. Kallin, and M. B. Hastings,
\newblock Phys. Rev. B {\bf 82}, 100409(R) (2010).

\bibitem{Kallin11165134}A. B. Kallin, M. B. Hastings, R. G. Melko, and R. R. P. Singh,
\newblock Phys. Rev, B {\bf 84}, 165134 (2011).

\bibitem{Hastings10157201} M. B. Hastings, I. Gonz\'{a}lez, A. B. Kallin, and R. G. Melko,
\newblock Phys. Rev. Lett. {\bf 104}, 157201 (2010).

\bibitem{Singh1275106} R. R. P. Singh, R. G. Melko, and J. Oitmaa,
\newblock Phys. Rev. B {\bf 86}, 075106 (2012).

\bibitem{Verstraete0487201} F. Verstraete, M. A. Mart\'{\i}n-Delgado, and J. I. Cirac,
\newblock Phys. Rev. Lett. {\bf 92}, 087201 (2004).

\bibitem{Popp0542306} M. Popp, F. Verstraete, M. A. Mart\'{\i}n-Delgado, and J. I. Cirac,
\newblock Phys. Rev. A. {\bf 71}, 042306 (2005).

\bibitem{Zanardi0242101} P. Zanardi,
\newblock Phys. Rev. A {\bf 65}, 042101 (2002).

\bibitem{Wang0422301} J. Wang and S. Kais,
\newblock Phys. Rev. A. {\bf 70}, 022301 (2004).

\bibitem{Gu0486402} S.-J. Gu, S.-Sa. Deng, Y.-Q. Li, and H.-Q. Lin,
\newblock Phys. Rev. Lett. {\bf 93}, 086402 (2004).

\bibitem{Johannesson07935} H. Johannesson and D. Larsson,
\newblock Low Temperature Physics {\bf 33}, 935 (2007).

\bibitem{Larsson05196406} D. Larsson and H. Johannesson,
\newblock Phys. Rev. Lett. {\bf 95}, 196406 (2005).

\bibitem{Larsson0642320} D. Larsson and H. Johannesson,
\newblock Phys. Rev. A. {\bf 73}, 042320 (2006).

\bibitem{Chan08345217} W.-L. Chan and S.-J. Gu,
\newblock J. Phys.: Condens. Matter {\bf 20}, 345217 (2008).

\bibitem{Mott49416} N. F. Mott,
\newblock Proc. Phys. Soc. London, Sect. B {\bf 62}, 416 (1949).

\bibitem{McWhan703734} D. B. McWhan and J. P. Remeika,
\newblock Phys. Rev. B {\bf 2}, 3734 (1970);
 D. B. McWhan {\it et al}., ibid.
\newblock {\bf 7}, 1920 (1973).

\bibitem{Greiner0239} M. Greiner, O. Mandel, T. Esslinger, T. W. H$\ddot{a}$nsch, and I. Bloch,
\newblock Nature (London) {\bf 415}, 39 (2002).

\bibitem{Gebhard1} F. Gebhard,
\newblock {\em The Mott Metal-Insulator Transition},
\newblock Springer-Verlag Berlin Heidelberg, 1997.

\bibitem{Byczuk1287004} K. Byczuk, Jan Kunes, W. Hofstetter, and D. Vollhardt,
\newblock Phys. Rev. Lett. {\bf 108}, 087004 (2012); {\bf 108}, 189902 (2012).

\bibitem{Metzner89324} W. Metzner and D. Vollhardt,
\newblock, Phys. Rev. Lett. {\bf 62}, 324 (1989).

\bibitem{Georges9613} A. Georges, G. Kotliar, W. Krauth, and M. J. Rozenberg,
\newblock Rev. Mod. Phys. {\bf 68}, 13 (1996).

\bibitem{Gottlieb07815} A. D. Gottlieb and N. J. Mauser,
\newblock Int. J. Quantum. Inform. {\bf 5}, 815 (2007).

\bibitem{Bulla99136} R. Bulla,
\newblock Phys. Rev. Lett. {\bf 83}, 136 (1999).

\bibitem{Tong01235109} N. H. Tong, S. Q. Shen, and F. C. Pu,
\newblock Phys. Rev. B {\bf 64}, 235109 (2001).

\bibitem{Caffarel941545} M. Caffarel and W. Krauth,
\newblock Phys. Rev. Lett. {\bf 72}, 1545 (1994).

\bibitem{Potthoff01165114} M. Potthoff,
\newblock Phys. Rev. B {\bf 64}, 165114 (2001);
J. Ortloff, M. Balzer, and M. Potthoff,
\newblock Eur. Phys. J. B {\bf 58}, 37 (2007).

\bibitem{Bulla00257} R. Bulla and M. Potthoff,
\newblock Eur. Phys. J. B {\bf 13}, 257 (2000).

\bibitem{Georges2} See Fig.34 of Ref.~\cite{Georges9613}.

\bibitem{Explain1} In Ref.~\cite{Potthoff01165114}, $z$ is
calculated via $z= 1/ \left[1- \partial \Sigma(\omega) / \partial
\omega |_{\omega=0} \right]$. Here, we calculate $z$ as the weight
of the qusi-particle poles in the Green's function. These two ways
have slight difference away from $V=0$.
\end{thebibliography}
\end{document}